# Mechanism and kinetics of bacteria-killing in a batch study in presence of Ag impregnated activated carbon

Vivekananda Bal

Department of Chemical Engineering, Indian Institute of Technology Bombay, Powai, Mumbai 400076, India

**Abstract:** A mathematical model based on classical species balance equation has been developed to explain the experimental observation on *E.Coli* cell-killing in a batch reactor and to extract information on cell-killing mechanism. Maximum likelihood optimization method has been used to obtain the best fit to the experimental data to extract ten unknown kinetic and thermodynamic parameters, which is otherwise practically impossible. Bacteria killing kinetics is found to be dominated by contact killing mechanism with approximately one order difference between contact-and bulk-killing kinetics. The order of Ag-bacteria interaction is highly non-linear and thus the interaction is highly complex in manner. Concentration of Ag on the outer surface of AC controls the kinetics of bacteria killing. Mechanism of release of Ag from the inner surface of pores does has minimal impact on the bacteria killing kinetics. Activity of silver seems to remain preserved after cell killing and the surface concentration of Ag used in the experiment is so high that bacteria killing kinetics is independent of whether Ag remains active or not.

## 1. Introduction

Drinking water contains pathogens like bacteria which cause water borne disease (WHO, 2014a; WHO, 2015). Removal of those bacteria from drinking water is a challenging task. One of the widely used methods of removal of these pathogens is allowing the contaminated water to pass through a packed bed filled with material having antibacterial property. One such material is Ag impregnated granular activated carbon (GAC) possessing a higher surface to volume ratio due to the highly porous nature of the carbon. Though there has been a plethora of experimental works to understand the bacteria removal efficiency of such a system, there are not many mathematical models to understand the mechanism of killing of bacteria in these systems.

The effect of Ag nanoparticles impregnated AC in removing bacteria from drinking water was studied by many researchers worldwide (Acevedo et al., 2014; Biswas and Bandyopadhyaya, 2016; Pal et al., 2006; Park et al., 2009; Srinivasan and Bandyopadhyaya, 2013; Zhao et al., 2013). The first study (Pal et al., 2006) of the antibacterial activity of Ag modified AC found that more than 90% bacteria is killed within 60 minutes and leaching of Ag is low close to 40 ppb. Antibacterial activity of Ag was found to be due to Ag mediated production of reactive oxygen (Park et al., 2009). Impregnated Ag particles were found to be present on the outer surface as well as inner surface of the AC particles, and impregnation of silver in the particle form reduces the release of silver into water (Zhao et al., 2013). Zhao et al., 2013 reported a Ag release of more than 25% of the deposited amount within 5-6 days. Acevedo et al., 2014 also fund that Ag nanoparticles were present on the outer as well as inner surface of AC and treatment of AC with KOH before Ag nanoparticle impregnation gives higher nanoparticle density and higher antibacterial activity with complete killing of bacteria within 15 minutes. However, they did not report any Ag release statistics. Later, it was found that plasma treatment of AC allows the deposition of Ag particles mostly on the outer surface of AC (Biswas and Bandyopadhyaya, 2016; Srinivasan and Bandyopadhyaya, 2013) and that reduces the Ag release to 4wt% within 10 days (Biswas and Bandyopadhyaya, 2016). This suggests that the Ag release and bacteria killing efficiency are the two challenging problems and there have been quite a few experimental works in addressing those problems. Understanding whether the bacteria killing mechanism is via contact killing or leaching of Ag followed by bulk killing and obtaining killing kinetics will help experimentalists in designing the experiments to obtain better killing efficiency. Determination of killing kinetics or mechanism experimentally can be a challenging task. In this regard, mathematical models can be useful in obtaining information on killing mechanism as well as kinetics.

In order to understand the bacterial random motion and chemotaxis in confined geometry, several discrete and continuum models have been developed over the years (Berg, 1983; Keller and Segel, 1971). Most popular one was the continuum model developed by Keller and Segel, 1971 based on the classical species balance equation. Later, many improved models have been proposed by (Dahlquist et al., 1976; Ford and

Cummings, 1992; Ford and Lauffenburger, 1991; Rivero et al., 1989)incorporating some modifications in the existing models to find the diffusion and chemotactic coefficient. Though modeling of packed bed reactor has been widely used (Cordero-Lanzac et al., 2019; Fogler, 2005; Zhu et al., 2020)to extract reaction kinetics and concentration profile of species in many different problems across many disciplines and is a classical method, there are not many rigorous and robust mathematical models to understand the mechanism of killing and to extract the killing kinetics.

Thus, it is clear that there have been many attempts either to understand the bacterial transport mechanism in a restricted environment or to extract chemical reaction kinetics in porous media or on catalyst particle surface, but no attempts have been made to understand the mechanism of killing of bacteria or the kinetics, specifically in the environment where antibacterial agent is present at the outer surface as well as in the inner surface of the complex tortuous pores of the AC. This is mainly because of the difficulty associated with performing experiment or the unavailability of experimental tools to track the killing mechanism. There are many theories on the mechanism of interaction of silver with bacteria. As per most of the literatures, silver penetrates the cell wall of the bacteria, binds with sulphur and phosphorous containing proteins, and thus prevent the cell division (Park et al., 2009; Yin et al., 2020). Therefore, a knowledge of the amount of Ag required to kill one CFU of bacteria and the knowledge of bacteria availability after killing are extremely important to design an optimum and efficient Ag-AC composite. In this work, to address above questions, a rigorous and robust mathematical model based on simple species balance equations has been presented by including in the model, the effects of (i) Ag impregnation in both inner and outer surface of AC, (ii) leaching of Ag into the bulk water from both outer and inner surface of AC, (iii) leach killing kinetics, and (iv) contact killing kinetics.

## 2. Experiment

Experimental data presented in this article was obtained from different sources (Biswas and Bandyopadhyaya, 2016; Pal et al., 2006; Zhao et al., 2013). Experimental procedure was described in detail in their work. Here we present a brief description of the experimental procedure for convenience of the readers. Ag nanoparticles were synthesized from AgNO$_3$ in presence of trisodium citrate dihydrate under UV light exposure. AC particles of size ranging from 420-840 μm was used as a supporting medium for Ag. AC pore size distribution was measured using mercury porosimeter and pores are micro and mesoporous in nature. To impregnate the Ag particles on AC particles, AC was first functionalized using plasma treatment in presence of oxygen atmosphere and then crushed gently and sieved using $20 \times 40$ mesh sieve. Sieved AC particles were then added in the Ag nanoparticles suspension and stirred overnight at room temperature. Thereafter, the Ag deposited AC particles were separated by filtration and then dried at room temperature.

For studying antibacterial activity of Ag-Ac, *E.Coli* K12 was used as a model bacteria. For experiments, *E.Coli* culture was prepared at a concentration of $10^4$ CFU/mL. Ag-AC granules at a concentration of 2, 4, 6, and 8 mg/mL were added in the cell culture and incubated and maintained at 150 rpm in a rotating shaker. Samples were collected every 5 minutes for a period of 35 minutes, plated on agar, incubated for 24 h, and then number of colonies formed were counted.

In the present work, both the AC size and AC pore size distribution are assumed to have a Gaussian distribution.

## 3. Model development

As we discussed, AC possesses a complex tortuous porous structure. In the presence of AC particles, bacteria will reach the surface of the carbon granules by self-diffusion process and come in contact with Ag nanoparticles. Since the AC granules used in this study were of micro, and mesoporous in nature, pore size remains below 50 μm (Biswas and Bandyopadhyaya, 2016; Pal et al., 2006; Zhao et al., 2013). Bacteria, *E.Coli*. K12 used in this study has a size greater than 1 μm in length and half a micrometer in diameter. This will prevent the bacteria from entering the pores of AC. Thus, bacteria will interact with Ag present on the outer surface of AC and the Ag present in the bulk liquid. Rate of change of bacteria concentration in the bulk liquid phase is given by

$$\frac{\partial c_b}{\partial t} = -r_{lb} - r_{sb} \qquad (1)$$

Here, $c_b$ is the bulk concentration of bacteria. $r_{lb}$ is the rate of killing of bacteria in the liquid phase. $r_{lb} = k_l c_{Ag,b}^m c_b^n$, here $k_l$ is the rate constant for killing of bacteria in liquid phase and $c_{Ag,b}$ is the concentration of Ag in the bulk liquid phase. $m$ and $n$ are the orders of the reaction with respect to bulk Ag and bacteria, respectively. $r_{sb}$ is the rate of killing of bacteria in the liquid phase adjacent to the surface of AC. $r_{sb} = k_s c_{Ag,sfo}^{m'} c_b^n$, here $k_s$ is the rate constant for killing of bacteria on the surface of GAC, $m'$ is the order of surface reaction with respect to surface silver. $c_{Ag,sfo}$ is the concentration of outer surface Ag. Since the system is well mixed and assuming that there is no diffusion limitation adjacent to the particle surface, bulk concentration of bacteria will be same as that in the liquid phase adjacent to the solid surface. In developing this model, it is assumed that the bulk solution is well mixed due to continuous stirring of the mixture.

In the literature, there are mixed reports on the release of silver from the surface (Acevedo et al., 2014; Biswas and Bandyopadhyaya, 2016; Pal et al., 2006; Zhao et al., 2013). Some claim a controlled and low release of Ag within the permissible limit, whereas some others report a significant release of silver into the water. This sparks the debate regarding the mechanism of killing of bacteria, leach killing vs. contact killing. Another important aspect is the loading of Ag on AC. In some work, Ag loading is mostly inside the pores (Pal et al., 2006; Zhao et al., 2013), whereas others (Biswas and Bandyopadhyaya, 2016) claim a selective impregnation of Ag on the outer surface of AC.

Here, two situations have been considered,

$CaseA: Bacteria + jAg = dead\ bacteria + jAg$

when activity of Ag is not lost, and

$CaseB: Bacteria + jAg = dead\ bacteria$

when Ag activity is lost.

From the stoichiometry balance (Fogler, 2005), we get,

$$r_{lAg} = jr_{lb}$$

and

$$r_{sAg} = jr_{sb}$$

Here, $r_{iAg}$ is the rate of loss of Ag due to interaction with Ag in the bulk liquid and $r_{sAg}$ is the rate of loss of Ag due to interaction with Ag on the outer surface of AC.

The equation 1 needs tracking of Ag concentration in the bulk liquid phase, which is given by

$$\frac{dc_{Ag,b}}{dt} = \frac{1}{V} \sum_{i=1}^{N} 2\pi R_i^2 N_{pi} D_{Ag} \frac{dc_{Ag,pi}}{dz}\bigg|_{z=L/2}$$
$$+ \frac{A_{GAC,o} k_{Ag}(c_{Ag,s} - c_{Ag,b})}{V} - r_{lAg} \quad (2)$$

$$r_{lAg} = \begin{cases} 0, & for\ no\ loss\ of\ Ag\ activity, caseA \\ jk_l c_{Ag,b}^m c_b^n, & for\ loss\ of\ Ag\ activity\ after\ killing, caseB \end{cases}$$

To take into account, the effect of loss of activity of Ag after killing of bacteria, one needs the information on stoichiometry, which is practically impossible to determine. Thus, stoichiometry of the reaction is assumed to follow the ratio $bacteria: Ag = 1: j$. $j$ is a positive integer number and value of $j$ used in the simulation varies between 1 to 20. $V$ is the volume of the bacteria containing water, $N_{pi}$ is the total number of pores of specific size $i$ in the AC added in that experiment, $R_i$ is the radius of a pore. $D_{Ag}$ is the silver diffusion coefficient in the tortuous interconnected pores in the AC. $c_{Ag,s}$ is the solubility of the Ag in liquid phase. It also represents the maximum attainable concentration of Ag in liquid phase adjacent to the solid surface due to dissolution of Ag coated on AC. $c_{Ag,pi}$ is the concentration of Ag inside pore. AC used in this study has a Gaussian pore size distribution. $A_{GAC,o}$ is the total outer surface area of the AC particles used in each experiment and $k_{Ag}$ represent the rate constant/mass transfer coefficient for release/dissolution of Ag into the liquid phase from the outer surface of AC.

Ag present in the bulk liquid comes from dissolution of Ag/Ag nanoparticles present on the outer surface of the AC particles and inside the pores on the AC. Thus, there is diffusion transport of Ag from the inside of the pores/AC particles towards the pore mouth/AC particle surface. $D_{Ag} \frac{dc_{Ag,pi}}{dz}\big|_{z=L/2}$ is the diffusion flux of Ag at the mouth of pore $i$ and the corresponding species balance equation for the transport (Fogler, 2005) of the Ag inside the pore $i$ is given as

$$\frac{\partial c_{Ag,pi}}{\partial t} = D_{Ag}\left[\frac{1}{r}\frac{\partial}{\partial r}\left(r\frac{\partial c_{Ag,pi}}{\partial r}\right) + \frac{\partial^2 c_{Ag,pi}}{\partial z^2}\right] \quad (3)$$

Here, $r_{Ag}$ is the rate of loss of Ag from the liquid phase due to killing of bacteria. Here, $r_{Ag} = r_l$. It is assumed that after killing of bacteria, activity of silver is lost. In developing the above model for transport of Ag, it is assumed that pores are straight cylindrical channels and open to media in both ends. Initial and boundary condition for solving the eq. 3 is given below

$$IC: t = 0, c_{Ag} = 0$$

$$BC1: z = 0, c_{Ag} = c_{Ag}$$

$$BC2: z = L/2, \frac{dc_{Ag}}{dz} = 0$$

$$BC3: r = 0, \frac{dc}{dr} = 0$$

$$BC4: r = R, flux = k'_{Ag}(c_{Ag,s} - c_{Ag,pi})$$

Here, pores are assumed to be cylindrical in shape, $D_{Ag}$ is effective diffusion coefficient of bacteria in the pore, $r$ is the radial co-ordinate, $z$ is the axial co-ordinate, $k'_{Ag}$ is the mass transfer coefficient for Ag release from the pore walls and $L$ is the length of the pore. Pore radius and GAC granules are found to have a Gaussian distribution.

Change in the surface concentration of Ag is given by

$$A_{tot}\frac{dc_{Ag,sf}}{dt} = A_{GAC,o}\frac{dc_{Ag,sf,0}}{dt} + A_{GAC,in}\frac{dc_{Ag,sf,in}}{dt}$$
$$= -A_{GAC,o}k_{Ag}(c_{Ag,s} - c_{Ag,b}) - Vr_{sAg}$$
$$- \sum_{i=1}^{N}\int_{0}^{L} 2\pi R_i N_{pi} k'_{Ag}(c_{Ag,s} - c_{Ag,pi})\, dL \quad (4)$$

$$r_{sAg} = \begin{cases} 0, & for\ no\ loss\ of\ Ag\ activity, case\ A \\ jk_s c_{Ag,sfo}^{m'} c_b^n, & for\ loss\ of\ Ag\ activity\ after\ killing, case B \end{cases}$$

In actual GAC particles, pores are tortuous, interconnected, and are of varying cross-sectional area along the porous path. In addition, some pores are open at both ends and most of them have dead ends. To incorporate these effects, we define an effective diffusion coefficient $D_e$ in the porous GAC particle, a concentration of Ag $c_{Age}$ based on per unit volume of GAC particle, and rewrite the species balance equation (Fogler, 2005) for Ag in the GAC as given by

$$\frac{\partial c_{Age}}{\partial t} + \frac{\partial}{r^2 \partial}\left(-r^2 D_e \frac{\partial c_{Age}}{\partial r}\right) - S_v r_n = 0 \quad (5)$$

Here, $D_e = \frac{D_{Ag}\epsilon\sigma}{\tau}$ (Fogler, 2005), where $\epsilon$ is the porosity of the GAC particle, $\sigma$ is the constriction factor, and $\tau$ is the tortuosity. Pore constriction factor takes into account of the variation in the cross-sectional area that is normal to the diffusion path. Tortuosity is defined as the ratio of actual distance a molecule travels between two points to the shortest distance between those two points. $S_v$ is the surface area of AC per unit volume of AC. $r_n = k_n(c_{Ag,s} - c_{Ag,e})$, where, $k_n$ is the rate constant for release/dissolution of Ag from the GAC surface. Thus, the bulk concentration of Ag is tracked by the same equation 2 with modified flux term. Thus, the equation 2 becomes

$$\frac{dc_{Ag,b}}{dt} = \frac{1}{V}\sum_{i=1}^{N} 4\pi R_i^2 N_{di} D_e \frac{dc_{Age}}{dr}\bigg|_{r=d_i/2}$$
$$+ \frac{A_{GAC,o}k_{Ag}(c_{Ag,s} - c_{Ag,b})}{V} - r_{lAg} \quad (6)$$

Here,

$$r_{lAg} = \begin{cases} 0, & for\ no\ loss\ of\ Ag\ activity, caseA \\ jk_l c_{Ag,b}^{m} c_b^n, & for\ loss\ of\ Ag\ activity\ after\ killing, caseB \end{cases}$$

Here, $R_i$ and $d_i$ are the radius and diameter of GAC granules/particles of size class $i$, $N_{di}$ is the number of GAC granules of size class $i$.

Change in the surface concentration of Ag is given by

$$A_{tot}\frac{dc_{Ag,sf}}{dt} = A_{GAC,o}\frac{dc_{Ag,sf,0}}{dt} + A_{GAC,in}\frac{dc_{Ag,sf,in}}{dt}$$
$$= -A_{GAC,o}k_{Ag}(c_{Ag,s} - c_{Ag,b}) - Vr_{sAg}$$
$$- \sum_{i=1}^{N}\int_{0}^{d_i/2} 4\pi r_i^2 N_{di} k_n(c_{Ag,s} - c_{Ag,e})\, dr \quad (7)$$

Here,

$$r_{sAg} = \begin{cases} 0, & for\ no\ loss\ of\ Ag\ activity, caseA \\ jk_s c_{Ag,sf}^{m} c_b^n, & for\ loss\ of\ Ag\ activity\ after\ killing, caseB \end{cases}$$

Here, $r_i$ is the radius variable for GAC granule of size class $r_i$.

The experiment was performed in absence of either chemoattractant or chemorepellent, and therefore, model developed here does not include the effect of biased transport due to chemoattractant or chemorepellent. Also, since the experiment was carried out for a short period of time, formation of bacteria due to growth and loss of bacteria due to death is assumed to be insignificant to be considered in the model.

Therefore, there are two sets of equations. Set I consists of equations 1-4 and set II consists of equations 5-7 and equation 1. For the set 1, set of unknown parameters consists of $k_l, k_s, D, c_{Ag,s}, m, m', n, k_{Ag}, k'_{Ag}$ and for the set 2, set of unknown parameters consists of $k_l, k_s, c_{Ag,s}, D_e, m, m', n, k_{Ag}, k'_{Ag}, k_n$ in the model. Parameters were fitted to the experimental data and best fit was obtained using maximum likelihood estimation method (Gunawan et al., 2002), where objective function is given by

$$\min_{p} (C_e - C_s(p, p_0))^T V_\varepsilon^{-1} (C_e - C_s(p, p_0)) \quad (8)$$

Here, $p$ is the vector of parameter, $p_0$ is the vector of predetermined parameters, $C_e$ is the vector of experimental observations, $C_s$ is the vector of model predictions, and $V_\varepsilon$ is the matrix of experimental error variance.

## 4. Method of computation

Governing PDEs were discretized in spatial domain using finite difference method (backward and central difference scheme) to obtain a set of ODEs in time domain (Gupta, 2013; Patankar, 1980). ODEs were then discretized in time domain using implicit Euler scheme to obtain a set of nonlinear algebraic equations, which were then solved using Newton-Raphson method along with Gauss-Seidel and successive overrelaxation method. The value of overrelaxation factor is maintained between 0-1 (Gupta, 2013; Patankar, 1980). Around 50-120 grids were used in the radial direction depending on the size and 140-250 grids were used in the axial direction depending on the pore length, making a total of 7,000-30,000, which represents the optimum number of grids minimizing the total error (truncation error + round off error). Dense grids were used near the wall of a pore to capture the sharp variation of concentration, whereas in the central region, sparse grids were used. Mass balance for Ag as well as bacteria was checked in every time step and mass balance error was well below 2%. An adaptive time stepping method was used, where time step size was varied between $10^{-5}$ to $10^{-2}$ s. A relative tolerance value of $10^{-5}$ was used to obtain a converged value.

Model and simulation results were validated with the existing simulation results for species balance equation in the literature for the different systems. Species balance equation for Ag inside the pore eq 4 was validated with the model in presented in textbook (Fogler, 2005) for the conversion of propylene oxide to the propylene glycol. Similarly, model eq. 6 was validated with model presented in the textbook (Fogler, 2005) for concentration profile in a spherical catalyst.

## 5. Results and discussion

A total of five data sets were taken from Biswas and Bandyopadhaya (Biswas and Bandyopadhyaya, 2016): four batch mode experimental data sets for bacteria killing and one batch date set for silver release. Two batch experimental data set for bacteria killing were taken from Pal et al. (Pal et al., 2006). Three experimental data sets were taken from Zhao et al. (Zhao et al., 2013): two batch experimental data sets for bacteria killing and one batch data set for Ag release. Thus, there are 10 unknown parameters and 10 experimental data sets to be fitted with. Parameters used in this simulation was taken from the respective articles (Biswas and Bandyopadhyaya, 2019, 2016; Pal et al., 2006; Zhao et al., 2013). Surface concentration of Ag is calculated from the BET/mercury porosimeter result and Ag loading data reported in the respective articles (Biswas and Bandyopadhyaya, 2019; Pal et al., 2006; Zhao et al., 2013). Distribution of silver between the inner and outer surface of AC is extremely important in this case as it varies with the methods used by different research groups to impregnate the Ag. In case of Ag-AC composite prepared by Pal et al. (Pal et al., 2006) and Zhao et al. (Zhao et al., 2013), Ag is assumed to be homogeneously distributed between inner and outer surface of AC, whereas for Biswas and Bandyopadhyaya (Biswas and Bandyopadhyaya, 2016), 50-80% of total Ag loading is assumed to be distributed homogeneously on the outer surface of AC and rest of the amount is assumed to be distributed homogeneously over the inner surface of the AC. In developing this model, it is assumed that Ag formed a continuous film on the AC surface. Since Ag particles used in their antibacterial activity study were larger in size compared to the micro-and mesopores, micro-and mesopores are omitted from Ag surface concentration calculation as it is less likely that those pores would be accessible to Ag particles.

Initial concentration of *E.Coli* in the antibacterial activity study was $10^4$ CFU/mL in case of Biswas and Bandyopadhyaya (Biswas and Bandyopadhyaya, 2016) and $10^7$ CFU/mL in case of Zhao et al. (Zhao et al., 2013) and Pal et al. (Pal et al., 2006). Similarly, initial volume of bacteria suspension and amount of Ag-AC used in the experiment were 100 mL and 2, 4, 6, 8 mg/mL, respectively, in case of Biswas and Bandyopadhyaya (Biswas and Bandyopadhyaya, 2016), 100 mL and 2 g, respectively, in case of Zhao et al. (Zhao et al., 2013), and 10 mL and 2 g, respectively, in case of Pal et al. (Pal et al., 2006). Initial loading of Ag on AC was 0.8 wt% in case of Biswas and Bandyopadhyaya (Biswas and Bandyopadhyaya, 2016), 1.65wt% in case of Zhao et al. (Zhao et al., 2013), and 0.2 wt% in case of Pal et al. (Pal et al., 2006).

Fitted simulation results shown in **Figs.2-4** are for the case A, where activity of Ag remains preserved even after interaction with bacteria cell. The corresponding fitted parameters are shown in **Table 1** for two sets of models set I and set II and for two different cases case

A and case B. It is observed that model simulation results explain the experimental variation very well in all cases. In case of **Fig.2**, faster killing of cell is achieved in case of batch study with higher Ag-AC concentration, which is obvious, as higher Ag-AC concentration increases the outer surface area density available for surface/contact killing per unit volume of the system. Both 6 and 8 mg/mL concentration shows the complete killing within 15 minutes suggesting that outer surface area limitation to the contact killing is not present in higher concertation. At low concertation of Ag-AC, low outer surface area density per unit volume of the system available for contact killing increases the killing time due to crowding of bacteria. Thus, killing time is higher in case of 2 mg/mL system as compared to the system with 4 mg/mL. Zhao et al. (Zhao et al., 2013; **Fig. 4**) found a time scale of 25 minutes for complete killing of bacteria for an Ag loading of 1.65wt%, which is twice than that used by Biswas and Bandyopadhyaya (Biswas and Bandyopadhyaya, 2016) for complete killing within similar time scale. This is probably because of the low surface concentration of Ag on the outer surface of AC in case of Zhao et al. (Zhao et al., 2013) as their method of Ag impregnation is non-selective. In case of Pal et al. (Pal et al., 2006; **Fig. 4**), time scale of complete killing is much higher than others because of the low surface Ag loading due to non-selective impregantion.

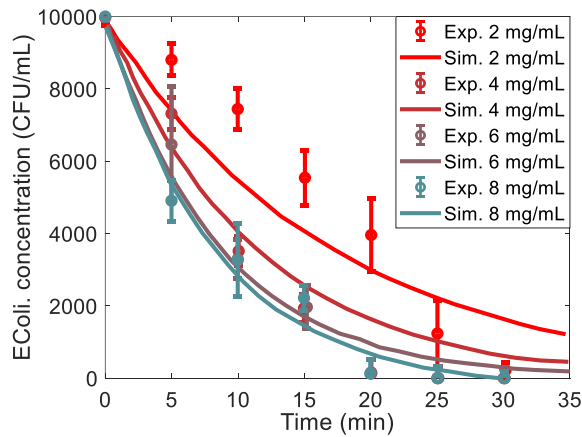

**Fig. 2:** Comparison of simulation results with the batch experimental date of Biswas and Bandyopadhyaya, 2016 for variation of *E.Coli* concentration as a function of time. Fitted parameters are listed in table 1. All parameters used in the simulation are taken from Biswas and Bandyopadhyaya, 2016.

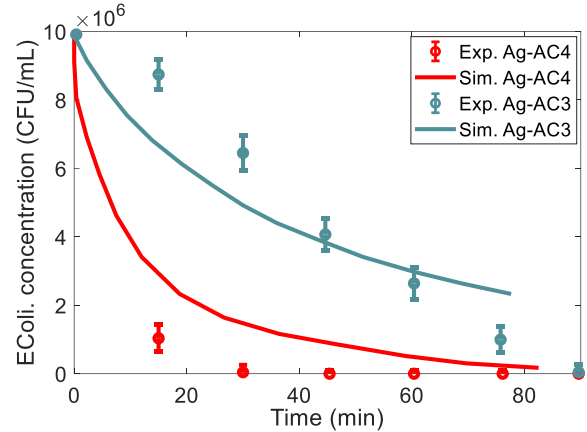

**Fig. 3:** Comparison of simulation results with the batch experimental date of Zhao et al., 2013 for variation of *E.Coli* concentration as a function of time. Fitted parameters are listed in table 1. All parameters used in the simulation are taken from Zhao et al., 2013.

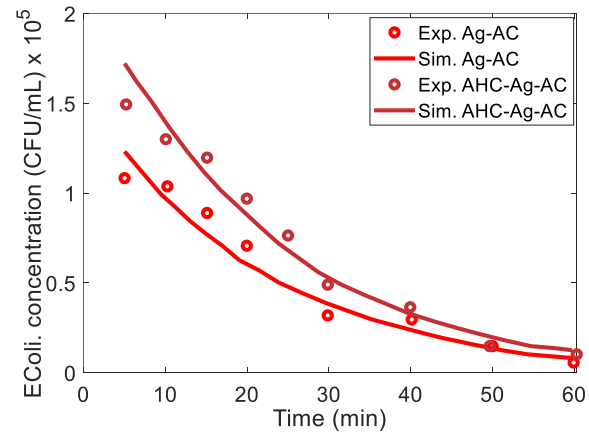

**Fig. 4:** Comparison of simulation results with the batch experimental date of Pal et al., 2006 for variation of *E.Coli* concentration as a function of time. Fitted parameters are listed in table 1. All parameters used in the simulation are taken from Pal et al., 2006.

**Figs. 5-6** Show the comparison of simulation results with the experimental data for Ag release. In both cases Ag release remains low even after a week, which has been captured by the model very well. Release rate or the total amount released is completely driven the driving force for dissolution, which is completely determined by solubility. Since the Ag solubility is low as shown in **table 1**, driving force is also low and this prevents the dissolution of more and more Ag as time progresses and Ag concentration saturates in the liquid.

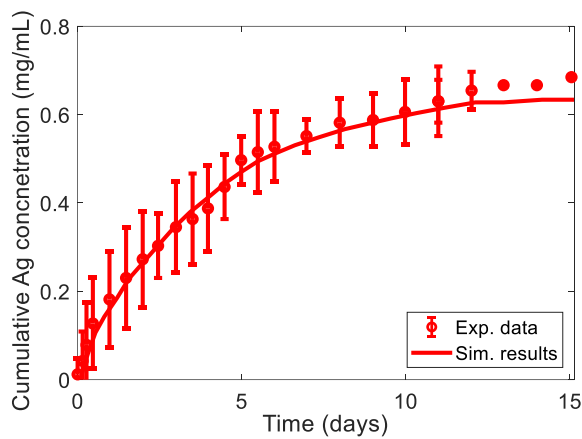

**Fig. 5:** Comparison of simulation results with the batch experimental data of Biswas and Baldyopadhyaya et al., 2016 for Ag release. Fitted parameters are shown in **table 1**.

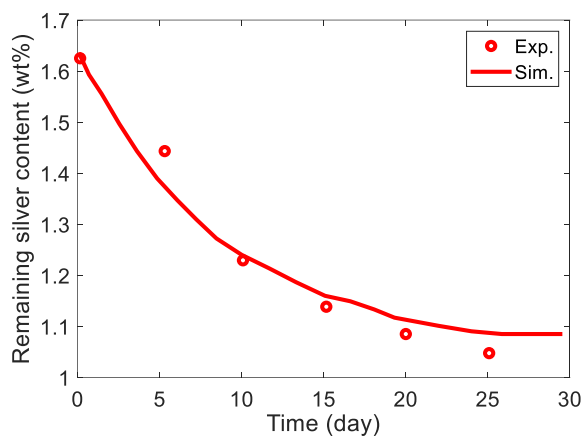

**Fig.6:** Comparison of simulation results with batch experimental data of Zhao et al., 2013 for Ag release. Fitted parameters are shown in **table 1.**

From the **table 1**, it is observed that value of rate constants for bulk-killing of bacteria is more than one order low as compared to that for surface/contact-killing. Thus, the bacteria killing mechanism is dominated by the contact killing. One possible reason for this is that the mechanism of bacteria-surface interaction in contact killing is different from that in the bulk killing. Another possible reason may be that bacteria can interact with more Ag atoms at the surface than in the bulk. Killing kinetics in both surface and bulk killing is not truly first order with respect to either Ag or bacteria as opposed to the claims made by most of the previous works (Biswas and Bandyopadhyaya, 2019; Pal et al., 2006; Zhao et al., 2013) in this area. Nonlinearity of the reaction order suggests that the interaction between bacteria and silver is quite complex.

There are many theories on the mechanism of interaction of silver with bacteria. According to most of the literature (Park et al., 2009), silver penetrates the cell wall of the bacteria, binds with sulphur and phosphorous containing proteins, and thus prevent the cell division. In order to calculate the amount of Ag required to kill one CFU of bacteria and to understand whether Ag remains available after killing bacteria (i.e., active Ag or inactive Ag), simulation was carried out with different stoichiometry ($j$) of the cell-bacteria interaction and without the loss of Ag activity as shown in the section 4. Overall, the calculated rate constants and order of the Ag-AC interactions do not show any significant variation from each other. Thus, it is impossible to draw any conclusion about the Ag activity. This is probably because the amount of Ag used in the experiments is so high that even if there is a loss of small amount of Ag due to inactivity, it does not affect the killing rate. It may be that the variation in stoichiometric amount of Ag from 1 to 100 is not sufficient and probably an even higher order of variation is necessary to see any significant variation in killing rate between case A and case B. However, numerical convergence problems above a stoichiometric value of 100 made it impossible to explore higher stoichiometric interactions. Little variation in rate constants and order of the reaction on variation of stoichiometric amount of Ag from 1 to 100 may also suggest that the Ag-AC interaction pathways or mechanism of interaction remains same and is independent of Ag involved in the interaction. Both the models as represented by set I and set II does not show a significant difference in parameter values. This indicates that the mechanism of Ag release from the inner pore surface of AC does not impact the killing kinetics.

The value of Ag diffusion coefficient obtained from this analysis is close to that reported in literature for other similar metal atoms or ions or molecules (Bilbao et al., 2016; Green and Southard, 2018). At least the calculated value is within the same order of magnitude. Equivalent diffusivity calculated from the second mode is slightly lower as compared to that obtained from the first model. This is because the second model considers the tortuous nature of the actual pores in the AC. Literature reports that Ag is sparingly soluble. The solubility value obtained in this analysis seems to be reasonable for sparingly soluble metal ions or atoms (Green and Southard, 2018). Mass transfer coefficient calculated in this analysis is within the range of that reported in literature for other materials or small molecules (Bilbao et al., 2016; Green and Southard, 2018).

Table 1: Parameters values obtained after fitting the experimental data.

| | $k_l \times 10^{-5}$ $\left(\dfrac{CFU\ mole^{-(m+n)}\ cm^{3(m+n-1)}}{s^{-1}}\right)$ | $k_s \times 10^{-6}$ $\left(\dfrac{CFU\ mole^{-(m'+n)}\ cm^{3(m'+n-1)}}{s^{-1}}\right)$ | $m$ | $n$ | $m'$ | $D \times 10^5$ or $D_e \times 10^5$ $\left(\dfrac{cm^2}{s}\right)$ | $c_{Ag,s}$ (ppb) | $k_{Ag}$ $\left(\dfrac{cm}{s}\right)$ | $k'_{Ag}$ $\left(\dfrac{cm}{s}\right)$ | $k_n$ $\left(\dfrac{cm}{s}\right)$ |
|---|---|---|---|---|---|---|---|---|---|---|
| Set I Case A | 2.351 | 1.721 | 1.354 | 1.172 | 1.321 | 1.648 | 915 | 6.113 | 9.27 | |
| Set II Case A | 2.290 | 1.652 | 1.271 | 1.032 | 1.219 | 0.852 | 857 | 6.912 | | 5.922 |
| Set I Case B $j = 1$ | 2.437 | 2.186 | 1.293 | 1.195 | 1.246 | 1.232 | 792 | 6.311 | 9.18 | |
| Set II Case B $j = 1$ | 2.189 | 1.897 | 1.281 | 1.157 | 1.278 | 0.759 | 771 | 6.252 | | 5.680 |
| Set I Case B $j = 100$ | 2.731 | 2.573 | 1.317 | 1.213 | 1.281 | 1.395 | 823 | 5.842 | 8.87 | |
| Set II Case B $j = 100$ | 2.335 | 2.023 | 1.239 | 1.162 | 1.253 | 0.829 | 895 | 6.248 | | 6.833 |

## 6. Conclusion

Model developed based on species balance equation including the effect of distribution of Ag on the outer as well as inner surface, with and without the loss of Ag activity, stoichiometry in Ag-bacteria interaction and Ag-bacteria interaction on the surface as well in the bulk nicely captured the experimental data for variation in bacteria killing and Ag release over time. It is found that the mechanism of release of Ag from the internal pores of AC does not influence the bacteria killing kinetics. Interestingly, it is found that the contact killing of bacteria dominates over the killing in bulk liquid (leach killing). Within the limit of this model, it is also found that the activity of Ag remains preserved after interaction with bacteria (i.e., Ag becomes available after interaction) and Ag-bacteria interaction is not limited by amount of Ag. Amount of Ag interacting with a bacteria CFU does not significantly influence the killing kinetics. Thankfully, low solubility of Ag prevents the release/dissolution of surface Ag. Model suggests that the concentration of Ag on the outer surface controls the overall killing kinetics. Mathematical model presented herein can be further improved by incorporating the effect of bacteria-Ag surface interaction mechanism, surface diffusion, resistance offered by liquid film around the AC surface to the bacteria diffusion, actual spatial distribution of Ag-AC composite particles in the liquid, mixing in the system and discrete Ag particles on AC surface instead of continuous film assumption.

## Acknowledgements

Author sincerely acknowledges the department of chemical engineering at Indian Institute of Technology Bombay for making this work to happen.